\documentclass[ twocolumn,
aps,prd,   
               preprintnumbers,numbers,sort&compress,nofootinbib,
                            showpacs,
               colorlinks,
               linkcolor=blue,   
               citecolor=blue]{revtex4-1}
   \newcommand{\exclude}[1]{}

\usepackage{graphicx,amsmath,amssymb,bm}
\usepackage{psfrag}
\usepackage{feynmp}
\usepackage{hyperref}
\usepackage{enumitem}

\newcommand{\beq}{\begin{equation}}
\newcommand{\eeq}{\end{equation}}
\newcommand{\be}{\begin{eqnarray}}
\newcommand{\ee}{\end{eqnarray}}

\def\+{\dagger}

\def\ra{\rangle}
\def\<{\langle}
\def\>{\rangle}

\newcommand{\Lqcd}{\Lambda_{\mathrm{QCD}}}
\newcommand{\Lbar}{\Lambda_{\overline{\mathrm{QCD}}}}
\newcommand{\qcd}{{\overline{\mathrm{QCD}}}}

\begin{document}

%\begin{frontmatter}

\title{Cosmological perturbations in $\qcd$- inflation. Estimates confronting  the  observations, including BICEP2. }

\author{Ariel R. Zhitnitsky} 
 \affiliation{Department of Physics \& Astronomy, University of British Columbia, Vancouver, B.C. V6T 1Z1, Canada}

%\date{\today}

\begin{abstract}
We discuss a new scenario for early cosmology, with the inflationary de Sitter phase dynamically emergent. 
This genuine quantum effect occurs as a result of the dynamics of topologically nontrivial sectors in a (conjectured) strongly coupled QCD- like gauge theory with the scale $\Lbar \sim 10^{17}$ GeV in an expanding universe. 
The inflaton in this $\qcd$-inflation framework is given by an auxiliary non-propagating field, similar to an effective field known to emerge in topologically ordered condensed matter systems.
The number of e-folds in the $\qcd$-inflation framework is determined by the gauge coupling constant at the moment of inflation, and estimated as $N_{\rm inf}\sim \alpha^{-2}(H_0)\sim 10^2$.
We analyze the equation of state in this framework at the end of inflation in terms of the gauge dynamics and confront our estimates with  observations.
We make predictions for the tensor tilt $n_t  \simeq -0.02$, the running of the tensor tilt $\alpha_t= {\partial n_t}/{\partial \ln k}\sim \alpha^4(H_0)\sim 10^{-4}$, and the running for the spectral index $\alpha_s= {\partial n_s}/{\partial \ln k}\sim \alpha^4(H_0)\sim 10^{-4}$ in terms of the same gauge coupling constant, $\alpha(H_0)$, which is fixed in our framework by recent BICEP2 measurements of the tensor fraction $r \simeq 0.2$. 
\end{abstract}

\maketitle

%\end{frontmatter}
\section{Introduction. Motivation.}
The main motivation for the present studies is the recent detection of the primordial gravitational waves by the BICEP2 collaboration\cite{BICEP2}.
These measurements ignited enormous excitement with many profound implications for the dynamics of the inflationary universe\cite{inflation}.
The corresponding studies, in most cases, are formulated in terms of constraints on the inflaton potential $V(\Phi)$ in a variety of models in which the inflaton field $\Phi$ is a real dynamical propagating degree of freedom which must satisfy a number of restrictions, see e.g. \cite{implications}.
In this letter we advocate a fundamentally new paradigm wherein the inflationary de Sitter behaviour emerges dynamically without any fundamental fields such as an inflaton $\Phi$.
In other words, the scale parameter ${\rm{a}} (t)$ and the equation of state takes the approximate form
\be	\label{a}
  {\rm{a}}(t)\sim \exp (Ht), ~~~ \epsilon\approx  -p
\ee
without any new propagating degrees of freedom as proposed in \cite{Zhitnitsky:2013pna}.

The new crucial element (which was not available at the time of publication of the original paper \cite{Zhitnitsky:2013pna}) is the measurements of the tensor fraction $r \simeq 0.2$ by BICEP2.
This measurement unambiguously fixes the Hubble constant $H_0\simeq 10^{14}$ GeV at the time of inflation.
Since the Hubble constant $H_0$ is uniquely expressed in terms of a new scale $\Lbar$ of a conjectured strongly coupled gauge theory (the so-called $\qcd$), this new scale is also unambiguously fixed, and is equal to $\Lbar \simeq 10^{17}$ GeV, which is not far from the Planck scale $M_{\rm PL}\simeq 2.4 \cdot 10^{18}$ GeV.
The fixing of this fundamental new scale allows us to make some specific predictions for a large number of observables such as the tensor tilt $n_t  \simeq -0.02$, the running of the tensor tilt $\alpha_t= {\partial n_t}/{\partial \ln k}$, and the running for the spectral index $\alpha_s= {\partial n_s}/{\partial \ln k}$ within this framework which we call the $\qcd$-inflation. 

The new paradigm advocated in \cite{Zhitnitsky:2013pna} is based on a fundamentally novel view on the nature and origin of the inflaton field which is drastically different from the conventional viewpoint that the inflaton is a dynamical local field.
In this new framework the inflation is a {\it genuine quantum effect} in which the role of the inflaton is played by an auxiliary topological field.
A similar field, for example, is known to emerge in the description of a topologically ordered condensed matter (CM) system realized in nature.
This field does not propagate, does not have a canonical kinetic term, as the sole role of the auxiliary field is to effectively describe the dynamics of the topological sectors of a gauge theory which are present in the system.
The corresponding physics is fundamentally indescribable in terms of any local propagating fields (such as $\Phi(x)$).
It might be instructive to get some intuitive picture for the $\qcd$-{\it inflaton} in this framework formulated in terms of a CM analogy as suggested in \cite{Zhitnitsky:2013pna}.
Such an intuitive picture is quite helpful in getting a rough idea about the nature of the inflaton in our framework.

Imagine that we study the Aharonov-Casher effect.
We insert an external charge into a superconductor in which the electric field is exponentially suppressed $\sim \exp(-r/\lambda)$ with $\lambda $ being the penetration depth.
Nevertheless, a neutral magnetic fluxon will be still sensitive to an inserted external charge at arbitrary large distances in spite of the screening of the physical field.
This genuine quantum effect is purely topological and non-local in nature and can be explained in terms of the dynamics of the gauge sectors which are responsible for the long range dynamics.
Imagine now that we study the same effect but in expanding universe.
The corresponding topological sectors will be modified due to the external background.
However, this modification can not be described in terms of any local dynamical fields, as there are not any propagating long range fields in the system since the physical electric field is screened.
For this simplified example, the dynamics of the inflaton corresponds to the effective description of the modification of topological sectors when the background changes. 
The effect is obviously non-local in nature as the Aharonov-Casher effect itself is a non-local phenomenon. 
    
One should emphasize that many relevant elements which are required for the inflationary phase have in fact been tested using the numerical lattice Monte Carlo simulations in strongly coupled QCD. 
However, in ref. \cite{Zhitnitsky:2013pna} we explained some deep physics questions related to the large distance behaviour in terms of a simplified version of QCD, the so-called ``deformed QCD" which is a weakly coupled gauge theory, but nevertheless preserves all the crucial elements of strongly interacting QCD, including confinement, nontrivial $\theta$ dependence, degeneracy of the topological sectors, etc.
In particular, the computation of the so-called ``strange energy''\footnote{This type of energy was coined as the ``strange energy" because it  can not be associated with any propagating degrees of freedom. This fundamentally new sort of energy can be in principle studied in tabletop experiments by measuring some specific corrections to the Casimir pressure, see  remarks and references in concluding section \ref{conclusion}.} which is the source for the behaviour (\ref{a}) has been explicitly performed in ref. \cite{Zhitnitsky:2013pna} in this simplified model. 

Our presentation is organized as follows.
In section \ref{review} we overview the basic results of ref. \cite{Zhitnitsky:2013pna}.
We explicitly formulate the assumptions which lead to the de Sitter behaviour (\ref{a}).
We emphasize that the regime (\ref{a}) would be the final destination of our Universe if interaction with standard model (SM) fields is switched off.
When the coupling is switched back on, the end of inflation is triggered precisely by this interaction which itself is unambiguously fixed by the triangle anomaly.
In section \ref{estimates} we use this anomalous coupling to express the observables such as number of e-folds $N_{\rm inf}$, the tensor fraction $r$, and other observables in terms of the parameters within the $\qcd$ framework, such as gauge coupling constant $\alpha(H_0)$ and $\Lbar$.
Section \ref{conclusion} is our Conclusion where we list few model-independent solid consequences of this new framework.
We also mention a possibility for testing the nature of the ``strange energy",  which is a key element of this framework, in a tabletop experiment by measuring some specific corrections to the Casimir vacuum energy in the Maxwell theory.

\section{The $\qcd$ -inflation. The basics.}	\label{review} 
The $\qcd$-inflation paradigm is based on three basic assumptions:\\
1. We assume there existence of a scaled up version of QCD (which is coined in ref. \cite{Zhitnitsky:2013pna} as $\qcd$) determined by the scale $\Lbar$.
It is not really a very new idea, rather a similar construction (though in a different context) has been suggested long ago for a different purpose and is known as technicolor.\\
2. We adopt the paradigm that the relevant definition of the energy in an expanding background, characterized by the parameter $H$, and which enters the Einstein equations, is the difference $\Delta E (H)\equiv \left[E(H)-E_{\mathrm{Mink}}\right]$, similar to the Casimir effect.
This element in our analysis is also not very new, and in fact in the present context such a definition for the vacuum energy was advocated in 1967 by Zeldovich \cite{Zeldovich:1967gd} for the first time; see \cite{Sola:2013gha} for review.\\
3. A novel element which was not widely discussed previously in the literature is an assumption that the ``strange'' vacuum energy (\ref{FLRW}), which can not be identified with any propagating degrees of freedom, receives the linear corrections $\sim H$ in the expending background, in apparent contradiction with conventional arguments that the corrections must be quadratic $\sim H^2$.
In other words, we assume that the expression for the vacuum energy in context of the Friedmann-Lema\^itre-Robertson-Walker (FLRW) universe takes the following form
\be	\label{FLRW}
  E_{\mathrm{FLRW}}(H)\sim \left[\Lbar^4+ H\Lbar^3+ {\cal{O}}(H^2)\right].
\ee
  
We refer to ref. \cite{Zhitnitsky:2013pna} for the detailed discussions and references on the physical meaning of eq. (\ref{FLRW}). 
However, we would like to make few important remarks regarding eq. (\ref{FLRW}).
The energy (\ref{FLRW}) has non-dispersive nature, i.e. it can not be expressed in terms of any propagating degrees of freedom\footnote{\label{top} In particular, this energy can be expressed in terms of the contact term in the topological susceptibility for the $\qcd$ theory, which has the ``wrong sign'', by which we mean the sign which is opposite to the dispersive contributions related to physical propagating degrees of freedom.
This ``wrong'' sign has been confirmed by numerous lattice numerical simulations, and in fact is a required feature for the resolution of the so-called $U(1)$ problem in QCD.
Furthermore, this energy vanishes above the phase transition, see \cite{Zhitnitsky:2013pna} for the details and references.
It may have profound cosmological consequences, see comment in Section \ref{conclusion}.}.
This feature is a simple reflection of the fact that the energy (\ref{FLRW}) is saturated by tunnelling events between physically identical but topologically distinct topological sectors $|k\ra$.
The physics of tunnelling processes and the corresponding generated energy can not be described in terms of a local dynamical field $\Phi(x)$, as the tunnelling between topologically distinct sectors is fundamentally non-local phenomenon.
The source of the linear term $\sim H$ in eq. (\ref{FLRW}) is the inherent non-locality of the large gauge transformation operator $\cal{T}$ which itself is the key element in the mechanism of generating the ``strange'' energy\footnote{The large gauge transformation operators is  defined as follows ${\cal{T}}|k\ra=|k+1\ra$.
The ``strange energy'' (\ref{FLRW}) is non-perturbative in nature as it is generated as a result of tunnelling events between $|k\ra$ and $|k+1\ra$ topological sectors.}. 

Furthermore, the energy (\ref{FLRW}) can not be expressed in terms of any local operators such as curvature, which would be a conventional structure to emerge when physical propagating degrees of freedom are integrated out in the background of the gravitational field.
This feature is similar to the well known property of a topologically ordered phase in condensed matter physics wherein an expectation value of a local operator does not characterize the system.
Instead, a system should be described in terms of some non-local variables.
In particular, in the simplified model considered in \cite{Zhitnitsky:2013pna} the correction $\sim H$ emerges as a result of the mixture of the auxiliary topological field (effectively describing the dynamics of the topological sectors) with gravity, see footnote \ref{auxiliary} with a short comment on the nature and origin of this auxiliary field. 
   
We strongly suspect that the crucial element related to the emergence of a linear correction $\sim H$ in this simplified model is the presence of a nontrivial holonomy in that model.
As is known, a non-trivial holonomy (similar, in structure, to the the Polyakov's loop) is a gauge invariant description of non-local gauge configurations\footnote{One should comment here that the dynamical generation of a nontrivial holonomy is likely to be a key element    leading to confinement in strongly coupled QCD, see e.g. a recent review \cite{Shuryak:2014gja}.
It is important to emphasize that a nontrivial holonomy was introduced into the system by means of a formulation on a torus with finite size $L$ as a  technical trick to properly account for the infrared physics.
In the limit $L\rightarrow \infty$, the system is effectively defined in an infinite space-time.
However, the topological features of the system related to a nontrivial holonomy still remain in the infinite volume limit. 
We interpret such behaviour of the system as a dynamical generation of a nontrivial holonomy in a strongly coupled QCD.}.
In principle, it may lead to linear $\sim H$ corrections in observables as the holonomy is represented by a linear order operator with respect to the potential, in contrast with a curvature which is characterized by a quadratic structure.  
   
In fact, such linear $\sim H$ effects in energy momentum tensor has been recently observed in Monte Carlo lattice studies \cite{Yamamoto:2014vda}. 
The numerical results \cite{Yamamoto:2014vda} strongly support our assumption about linear corrections  $\sim H$ in the energy momentum tensor.
In particular, the results \cite{Yamamoto:2014vda} confirm our formula (\ref{production}) on particle production in time dependent background in QCD.
We think that the numerical result \cite{Yamamoto:2014vda} is a consequence of the formulation of the system on a torus when a nontrivial holonomy can be dynamically generated, as we discussed above. 

Finally, the energy (\ref{FLRW}) vanishes above the $\qcd$ phase transition in the deconfined phase as this structure emerges only as a result of confinement in $\qcd$ theory.
This is again in huge contrast with conventional inflationary scenarios when the fundamental dynamical field $\Phi$ and the potential  $V(\Phi)$ always existed, before and after the inflation.  
To conclude, the property of non-locality which is an inherent feature of QCD may falsify the main assumption leading to a naive $H^2$ prediction.

With these three assumptions, the Universe had a period of inflationary (almost) de Sitter phase characterized by the behaviour (\ref{a}).
Indeed, the Friedman equation assumes the following form
\be	\label{friedman-infl}
  H^2= \frac{8\pi G}{3}\left(  \rho_{\mathrm{Inf}}+\rho_R\right)=\frac{8\pi G}{3}\left( \overline{\alpha} H\Lbar^3+\rho_R\right),  
\ee
where we identify $\rho_{\mathrm{Inf}}$ with $\Delta E (H)=\left[E(H) -E_{\mathrm{Mink}}\right]$ according to postulate 2 formulated above.
Furthermore, the corresponding energy density according to eq. (\ref{FLRW}) is given by $\rho_{\mathrm{Inf}}=\overline{\alpha}H\Lbar^3$.
In this expression $\overline{\alpha}$ is a dimensionless parameter of order of one.
This numerical coefficient is, in principle, computable in strongly coupled $\qcd$ from first principles.
In (\ref{friedman-infl}) we neglected higher order corrections $ {\cal{O}}(\Lbar^2 H^2)$ in the expansion (\ref{FLRW}) as $H\ll \Lbar$, see below.
The radiation component in eq. (\ref{friedman-infl}) scales as $\rho_R\sim {\rm a}^{-4}$ such that $\rho_{\mathrm{Inf}}$ starts to dominate the universe at some point when $H$ approaches the constant value $H_0$ estimated as follows
\be	\label{H_0}
  H_0\simeq \frac{8\pi G}{3} ( \overline{\alpha} \Lbar^3 )\simeq \frac{\overline{\alpha}}{3}\frac{ \Lbar^3}{M^2_{\rm PL}}, ~~ M^{-2}_{\rm PL}\equiv \sqrt{8\pi G}.
\ee
The constant $H_0$, which is unambiguously determined by the strongly coupled dimensional parameter $\Lbar$ corresponds to the inflationary (almost) de Sitter behaviour such that the equation of state (EoS) and parameter ${\rm a}(t)$
are:
\be	\label{infl-EoS}
  \omega \equiv \frac{p}{\rho}\simeq -1  , ~~  {\rm a}(t)\sim \exp (H_{0}t).  
\ee
The inflationary regime described by eqs. (\ref{H_0}) and  (\ref{infl-EoS}) would be the final destination of our Universe if the interaction of the $\qcd$ fields with SM particles were always switched off.
When the coupling is switched back on, the end of inflation is triggered precisely by this interaction which itself is unambiguously fixed by the triangle anomaly as we review below.

Before we explain the structure of the relevant interaction we want to make few comments. 
First, the physics responsible for the ``strange energy'' (\ref{FLRW}) which eventually leads to the de Sitter behaviour  (\ref{infl-EoS})   can not be formulated in terms of any physical degrees of freedom as we already mentioned.
However, the relevant physics can be formulated in terms of some auxiliary fields which exactly saturate pertinent correlation functions with the ``wrong sign'' and which eventually generate the ``strange energy'' (\ref{FLRW}).
These auxiliary fields are not mandatory fields, but instead play a supplementary role to simplify the analysis\footnote{ \label{auxiliary}In the weakly coupled ``deformed QCD'' the corresponding computations can be explicitly carried out where the auxiliary topological fields can be  expressed in terms of the original fields of the underlying gauge theory\cite{Zhitnitsky:2013hs}.
It is a matter of convenience to perform the computations of the ``strange energy'' in terms of auxiliary fields instead of explicit summation over positions and orientations of the monopoles-instantons describing the tunnelling events in semiclassical approximation
as it was originally computed in \cite{Thomas:2011ee}.
This technical trick to introduce some auxiliary fields describing the long range dynamics of an original strongly correlated system is, in fact, used to study the dynamics in many strongly correlated condensed matter systems.
A well-known example is th quantum Hall effect where the emergent (auxiliary) fields are described by the Chern-Simons effective Lagrangian.}  of the dynamics of the multiple tunnelling transitions between the distinct topological sectors $|k\ra$ in strongly coupled $\qcd$.
The only information which is needed for the future discussions is that the auxiliary field $b(x)$, saturating the ``strange energy'' (\ref{FLRW}), couples to the SM particles precisely in the same way as the axion $\theta$ couples to the gauge fields, see \cite{Zhitnitsky:2013pna} for the details. 
Furthermore, the $b(x)$ field has the same $2\pi$ periodic properties as the axion field.
The difference with the dynamical axion $\theta(x)$ is that the auxiliary field $b(x)$ does not have a conventional axion kinetic term. 
In other words, the coupling is \cite{Zhitnitsky:2013pna}
\be	\label{coup}
  {\cal L}_{b\gamma\gamma} (x)= \frac{\alpha (H_0)}{8\pi} N  Q^2 \left[ \theta- b(x)\right] \cdot F_{\mu\nu} \tilde F^{\mu\nu} (x) \, ,
\ee
where $\alpha(H_0)$ is the fine-structure constant measured during the period of inflation, $Q$ is the electric charge of a $\qcd$ quark, $N$ is the number of colours of the strongly coupled $\qcd$, and $F_{\mu\nu}$ is the usual electromagnetic field strength.
The coupling (\ref{coup}) is unambiguously fixed because the auxiliary $b(x)$ field always accompanies the so-called $\theta$ parameter in the specific combination $\left(\theta-b(x)\right)$ as explained in  \cite{Zhitnitsky:2013pna}, and describes the anomalous interaction of the topological auxiliary $b(x)$ field with $E\&M$ photons.
The coupling  of the $b(x)$ with other $E\&W$ gauge bosons can be unambiguously reconstructed as explained in \cite{Zhitnitsky:2013pna}, but we keep a single $E\&M$ field $F_{\mu\nu}$ to simplify the notations and emphasize on the crucial elements of the dynamics, related to the helical instability which trigers the end of inflation, see next section.
We take $\theta=0$ in eq. (\ref{coup}) as we do not intend to discuss in this work an interesting, but different, subject related to the dynamics of the physical axion field.  

As a result of these simplifications, in the numerical estimates in section \ref{estimates}, the coupling constant $\alpha(H_0)$ 
should be treated as an effective phenomenological parameter describing the dynamics of all gauge fields of the standard model during the inflation.  
 
\section{The end of inflation and the Cosmological perturbations}\label{estimates} 
The main goal of this section is to argue that the $\qcd$ inflation paradigm discussed in previous section is consistent with all presently available observations, including the recent measurements of the primordial gravity waves by BICEP2 collaboration\cite{BICEP2}. 
 
\subsection{The helical instability and the end of inflation }\label{instability}
It has been known for quite sometime that the structure of the interaction (\ref{coup}) in many respects has a unique and mathematically beautiful structure with a large number of very interesting features.
The most profound property which is crucial for our present analysis of the inflationary Universe is the observation that the topological term (\ref{coup}) along with the conventional Maxwell term $F_{\mu\nu}^2$ leads to an instability with respect to photon production in which   $\dot{b}(x)$ is non-vanishing. 
This is the so-called helical instability and has been studied in condensed matter literature \cite{Frohlich:2002fg} as well as in particle physics literature including some cosmological applications \cite{Joyce:1997uy}. 

In context of our studies, the closest system where the helical instability develops is the system of heavy ion collisions \cite{Akamatsu:2013pjd} wherein $\dot{b}(x)$ can be identified\footnote{The simplest way to demonstrate the correctness of this identification is to perform the path integral $U(1)_A$ chiral time-dependent transformation to rotate away the coupling (\ref{coup}).
The corresponding interaction reapers in the form of a non-vanishing axial chemical potential $\mu_5$, see Appendix B of ref.\cite{Zhitnitsky:2013pna} with details and references.} with the so-called axial chemical potential $\dot{b}(x)=\mu_5$.
One can explicitly demonstrate that the interaction (\ref{coup}) leads to the exponential growth of the low-energy modes with  $k\leq \frac{\alpha \mu_5}{\pi}$.
This growth signals that the instability of the system with respect to production of the real photons \cite{Akamatsu:2013pjd} develops.
It is also known that the fate of this instability is to reduce the axial chemical potential $\mu_5$ which was the source of this instability.
One should also comment here that parameter $\mu_5$ in heavy ion system is also not a dynamical field, but rather is an auxiliary fluctuating field which accounts for the dynamics of the topological sectors in QCD, similar to our case when $\dot{b}(x)$ describes the dynamics of the topological sectors in $\qcd$. 

This short detour into the nature of helical instability as a result of interaction (\ref{coup}) has direct relevance to our studies because the auxiliary field $b(x)$ entering eq.(\ref{coup}) exhibits all the features of parameter $\mu_5$ which was the crucial element in the analysis of the helical instability in heavy ion collisions.
Indeed, both these auxiliary fields originated from the same physics and they both describe the dynamics of the topological sectors in QCD and $\qcd$ correspondingly.
In physical terms these fields ($|\dot{b}(x)|\sim H$ and $\mu_5$) effectively account for the long range variation of the tunnelling processes as a result of some external influence of the backgrounds expressed in terms of $H^{-1}\gg \Lbar^{-1}$ for inflation and $\mu_5^{-1}\gg \Lqcd^{-1}$ for heavy ion collisions respectively
See some additional comments on this analogy in \cite{Zhitnitsky:2012im} and Appendix B of ref.\cite{Zhitnitsky:2013pna}.
One should also comment here that while the relevant analysis in strongly coupled QCD and $\qcd$ is the prerogative of the lattice numerical simulations, the corresponding questions can be addressed and analytically answered in a weakly coupled ``deformed QCD'' where the long range structure indeed emerges as a result of dynamics of the topological auxiliary field $b(x)$ with the axion quantum numbers, see \cite{Zhitnitsky:2013hs,Zhitnitsky:2012ej} with the details. 
  
The number of $e$-folds in the $\qcd$-inflation is determined by the time $\tau_{\rm inst}$ when the helical instability fully develops.  This is exactly the time scale where a large portion of the energy $\rho_{\mathrm{Inf}}$ related to the inflation from eq. (\ref{friedman-infl}) is transferred to SM light fields. 
The corresponding time scale for the heavy ion system is known \cite{Akamatsu:2013pjd} and it is given by $\tau_{\rm inst}^{-1} \sim \mu_5\alpha^2$.
For our system $\mu_5$ should be replaced by $|\dot{b}|\sim H$, and therefore we arrive at the following order of magnitude estimate for the number of $e$-folds $N_{\text{Inf}}$ in $\qcd$ inflationary paradigm, 
\be	\label{e-folds}
  \tau_{\rm inst}^{-1}\sim  {H_0\alpha^2(H_0)}, ~~~~\Longrightarrow ~~~N_{\text{Inf}}\sim \frac{1}{\alpha^2(H_0)},
\ee   
where number of $e$-folds $N_{\text{Inf}}$ is, by definition, the coefficient in front of $H_0^{-1}$ in the expression for the time scale $\tau_{\rm inst}$.
At this moment the inflation ceases as the dominant portion of the energy is already transferred to the light particles.
The key element of this $\qcd$ inflationary scenario is that the number of $e$-folds $N_{\text{Inf}}$ and the de Sitter behaviour (\ref{infl-EoS}) in this framework is determined by the gauge coupling constant $\alpha(H_0)$ rather than by dynamics of ad hoc inflaton $\Phi$ governed by some inflationary potential $V(\Phi)$. 

\subsection{Equation of State}		\label{EoS} 
The recent detection of the primordial gravitational waves by the BICEP2 collaboration \cite{BICEP2} implies that the Hubble constant $H_0\simeq 10^{14}$ GeV during inflation, since the tensor perturbations generated during the inflation are unambiguously expressed in terms of the Hubble parameter at the epoch of the horizon exit \cite{mukhanov}.
In our framework $H_0$ is uniquely fixed by eq. (\ref{H_0}).
Therefore, we can fix  our fundamental scale $\Lbar$ based on the BICEP2 measurements as follows
\be	\label{Lambda}
  \Lbar\simeq \sqrt[3]{\frac{3M^2_{\rm PL} H_0}{\bar{\alpha}}}\simeq 10^{17} {\rm GeV}.
\ee
This scale is slightly below the Planck scale, and therefore our treatment of the problem using the quantum field theory methods is still justified as the following hierarchy of scales emerges:
\be	\label{scale}
  M_{\rm PL} \gg  \Lbar \gg H_0.
\ee
  
Our next topic for discussions is the equation of state (EoS) during inflation as it enters the expressions for all relevant observables    such as the spectral index $n_s\simeq 0.96$, the tensor tilt $n_t$, the tensor fraction $r\simeq 0.2$, the running of the tensor tilt $\alpha_t= {\partial n_t}/{\partial \ln k}$, and the running for the spectral index $\alpha_s= {\partial n_s}/{\partial \ln k}$. 

In what follows we need the expression for $\omega\equiv p/\rho$ just before the inflation ends due to the development of the helical instability as described in section \ref{instability}.
However, for pedagogical reasons, we start our analysis for time $t$ soon after the inflation begins at $t_i$ but long before inflation ends at the moment $\tau_{\rm inst}$ with a fully developed helical instability.
In other words, we consider the time scale $(t-t_i)\ll \tau_{\rm inst}$, during the first few Hubble periods. 
In this case the number of produced particles per unit time per unit volume is determined by the coupling (\ref{coup}) and can be estimated as follows
\be	\label{production}
  \frac{dP}{dV dt}\sim \alpha^2(H_0) H_0\Lbar^3.
\ee
This formula describes (in physical terminology rather than in terms of the auxiliary $b(x)$ field) the production of real particles as a result of multiple tunnelling events between the topological sectors $|k\ra$ in the background of the gravitational field parameterized by the Hubble constant $H$.
The arguments supporting the linear dependence on $H$ are identical to those presented after eq.(\ref{FLRW}), and we shall not repeat them again.
In the case of Minkowski space-time when $H \rightarrow 0$ the tunnelling events are happening on a typical time scale $\Lbar$, but they obviously do not produce any particles, and the probability (\ref{production}) vanishes as it should. In fact, the lattice simulations \cite{Yamamoto:2014vda} also observe a linear dependence  on Hubble constant  $H_0$ for particle production rate, in complete agreement with our expression  (\ref{production}). 

The combination $H_0\Lbar^3$ entering eq. (\ref{production}) is nothing but the inflationary energy density (\ref{friedman-infl}). Therefore, one should expect some small correction $\sim \alpha^2(H_0)$ to the EoS given by eq. (\ref{infl-EoS}) as a result of the interaction with light particles (\ref{coup}).  

Our goal, however, is not the computation of the EoS at the beginning of inflation when ${(t-t_i)}\ll {\tau_{\rm inst}}$.
Rather, our goal is to compute the EoS at the very end of inflation when ${(t-t_i)}\simeq {\tau_{\rm inst}}$ just before the helical instability fully develops.
As we already mentioned the fate of this instability in heavy ion system is known: this instability reduces the axial chemical potential $\mu_5$ which was the source of this instability.
In our system $\dot{b}\sim H$ plays the same role as $\mu_5$ in heavy ion collisions, as we already mentioned.   
In our cosmological context such a flow of energy implies that the fate of instability is to reduce the inflationary Hubble constant $H$.
The corresponding inflationary energy which is proportional to the Hubble constant (\ref{friedman-infl}) will be transferred to the light particles during time $\tau_{\rm inst}$, which is precisely the destiny and fate of the reheating epoch.     
The corresponding time development of the helical instability which would provide this crucial information can, in principle, be carried out  from the first principles as all the relevant fundamental interactions are known.

In practice, however, this is a very technical numerical problem  which is yet to be solved.
Therefore, we choose a practical way to parametrize the EoS which properly reflects our understanding of the behaviour of the system while the helical instability develops.
We parametrize $\omega$ at the very end of inflation as
\be	\label{eos}
  \omega=\frac{p}{\rho} = -1+ c_2\alpha^2(H_0)\cdot e^{c_1\left[\frac{t-t_i}{\tau_{\rm inst}}-1\right]}, 
\ee
where the two numerical coefficients $c_1, c_2$ will be fixed using two measured observables: $r$ and $n_s-1$.
Our choice of the exponential function in time in (\ref{eos}) is based on our understanding of development of helical instability which  leads to a sharp end of the inflation.
Essentially we fix these two constants by fixing $\omega$ and its time derivative during the final moment of inflation as follows:
\be	\label{const}
  \left(\frac{p}{\rho}+1\right)_{t-t_i=\tau_{\rm inst}}=c_2\alpha^2(H_0)\\
  \frac{d \ln\left(\frac{p}{\rho}+1\right)}{dt}|_{t-t_i=\tau_{\rm inst}}=\frac{c_1}{\tau_{\rm inst}}.\nonumber
\ee
We must admit that the structure (\ref{eos}) is not based on solid theoretical computations. 
Therefore, we are not pretending to have made a solid prediction on behaviour of $\omega$ at the end of inflation.
Rather, our goal with eqs. (\ref{eos}) and (\ref{const}) is quite different; we want to argue that the available observational data can be easily accommodated within our framework of the $\qcd$-inflation.
In concluding section \ref{conclusion} we  list some  model independent consequences of the $\qcd$-paradigm.
These solid consequences should be contrasted with our model-dependent predictions, to be discussed in next subsection, and which are based on eqs. (\ref{eos}) and (\ref{const}).   

\subsection{Cosmological Perturbations}\label{observations}
We are now in position to fix the two free parameters from (\ref{eos}) using the measured values for the spectral index $n_s\simeq 0.96$ and  tensor fraction $r\simeq 0.2$ recently measured by BICEP2 collaboration\cite{BICEP2}.
We start with $r\simeq 0.2$.
In conventional inflationary scenarios based on the scalar potential the magnitude of $r$ is normally expressed in terms of the slow-roll parameters of the inflaton potential $V(\Phi)$, see e.g.\cite{implications}. 
In our framework we do not have scalar field, nor scalar potential $V(\Phi)$. Nevertheless, the EoS is perfectly defined for the system.
In fact, all observables can be directly expressed in terms of the EoS without even mentioning the potential $V(\Phi)$.
In particular, the expression for tensor fraction $r$ is given by \cite{mukhanov}
\be	\label{r}
  r\simeq 27 \left[|c_s|\left(\frac{p}{\rho}+1\right)\right]_{ k\simeq H{\rm a}},
\ee
where the so-called speed of sound in this expression is defined as $c_s^2=\partial{p}/\partial\rho$.
One should comment here that in the conventional description with inflation described in terms of the physical propagating degrees of freedom the parameter $c_s^2$ must be positive.
In such a conventional case a negative $c^2_s<0$ is considered as a signal of instability of the system.
There is no such requirement for our system as there are not any propagating degrees of freedom associated with this speed $c_s$.
In particular, in a pure de Sitter state $c_s^2=-1$ as one can see from (\ref{infl-EoS}), and it is obviously consistent with all fundamental theorems, see also an additional comment on $c_s$ in footnote 4 in ref.\cite{Zhitnitsky:2013pna}.  

One can easily show that in our case $|c_s|^2\simeq 1$ is very close to unity as $c_s^2$ receives very small corrections $\sim \alpha^2(H_0)$ which will be consistently ignored in our estimates. 
Comparing (\ref{r}) with (\ref{const}) we arrive at the condition which determines our parameter $c_2$,
\be	\label{r1}
  r\simeq 27  c_2\alpha^2(H_0)\simeq 0.2, 
\ee
where we used $r\simeq 0.2$ as measured by BICEP2\cite{BICEP2}.

Our next step is an analysis of the spectral index $n_s$ which is also known in terms of the EoS and it is given by\cite{mukhanov}
\be	\label{ns}
  n_s-1\simeq -3 \left(\frac{p}{\rho}+1\right)- \frac{1}{H}\frac{d \ln\left(\frac{p}{\rho}+1\right)}{dt}-\frac{1}{H}\frac{d \ln |c_s|}{dt}.
\ee
The last term in this formula is parametrically smaller $\sim \alpha^4$ than the  first two terms, and therefore, will be ignored.
As a result, we arrive at the condition which determines our parameter $c_1$,
\be	\label{ns1}
  n_s-1 &\simeq&  -3c_2\alpha^2(H_0)-\frac{c_1}{H_0\tau_{\rm inst}} \nonumber\\
        &\simeq& -\alpha^2(H_0) \left[3c_2+c_1\right]\simeq -0.04,
\ee
where we used $n_s\simeq 0.96$ as measured by PLANCK\cite{planck}.
Assuming $N_{\rm inf} \simeq 100$ and using estimates (\ref{e-folds}), (\ref{r1}), and (\ref{ns1}), we find the following set of parameters which approximately describe the observations: 
\be	\label{fit}
  c_2\simeq 0.74, ~ c_1\simeq 1.8, ~ \alpha(H_0)\simeq 0.1, ~N_{\rm inf} \simeq 100.
\ee
A few comments are in order.
First of all, it was not our goal to fit the data with perfect accuracy.
Such an analysis would be too premature at this point as a numerical understanding of the evolution of the helical instability (which determines the EoS) is yet to be fully developed.
Rather, our goal was to demonstrate that the $\qcd$-inflation, in principle, can easily accommodate the presently available observations.
Secondly, as we previously mentioned, the parameter $\alpha(H_0)$ entering (\ref{eos}) should be treated as an effective coupling constant at the scale $H_0$ which effectively accounts for other gauge (and matter) fields participating in the development of the helical instability.
As an oversimplified estimate\footnote{ this estimate is very primitive as strongly coupled gluons, while not directly coupled to $b(x)$ field, nevertheless may considerably influence the evolution of the helical instability due to the secondary interactions.} one can  approximate $\alpha(H_0)\simeq (\alpha_{EM}+3 \alpha_{EW}) \simeq 4/40\simeq 0.1$ as the number of gauge fields of the SM which directly couple to $b(x)$ is four.
While such an estimate is very primitive, it nevertheless agrees with eq. (\ref{fit}) obtained as a result of matching of EoS (\ref{eos}) with observations.

Now we are in a position to make some predictions for observables which have not been measured yet.
We start with the tensor tilt $n_t$.
The corresponding expression in terms of the EoS is known \cite{mukhanov} 
\be	\label{nt}
  n_t\simeq -3 \left(\frac{p}{\rho}+1\right)\simeq -3\alpha^2(H_0)c_2\simeq -0.02,
\ee
where for a numerical estimate we use (\ref{const}) and (\ref{fit}).
Our estimate is consistent with conventional predictions of slow roll inflation where $n_t\simeq -r/8$ \cite{implications}.
However, the prediction  (\ref{nt}) is in conflict with a proposal \cite{Smith:2014kka} that the Planck/BICEP2 tension can be lessened if the tensor tilt $n_t$ is very blue (positive and order of one).

Now we estimate the running of the spectral index $\alpha_s$.
The corresponding expression in terms of the EoS can be estimated as follows
\be	\label{ns2}
   \alpha_s&\simeq& \frac{\partial n_s}{\partial \ln k}\simeq \frac{1}{H}\frac{\partial n_s}{\partial t}\simeq \frac{-3}{H}\frac{\partial  \left(\frac{p}{\rho}+1\right)}{\partial t}\nonumber\\
   &\simeq& -3c_1c_2 \alpha^4(H_0)\simeq -4\cdot 10^{-4},
\ee
where we took into account that the differentiation with respect to $ \frac{\partial}{\partial \ln k}$ can be approximated as $\frac{\partial }{H\partial t}$ because $\ln k\sim \ln {\rm a}$.
We use expressions (\ref{ns}) and (\ref{eos}) for $n_s$ to complete the differentiation with respect to time.
Our estimate (\ref{ns2}) is again consistent with conventional predictions of slow roll inflation.
However, it is around 100 times smaller than the preferred value extracted from Planck and BICEP2 where $\alpha_s\approx -0.028$   \cite{Smith:2014kka}. 
  
Finally, we estimate the running of the spectral index $\alpha_t$.
The corresponding expression in terms of the EoS can be estimated in a similar manner with the result
\be	\label{nt2}
   \alpha_t&\simeq& \frac{\partial n_t}{\partial \ln k} \simeq  -3c_1c_2 \alpha^4(H_0)\simeq -4\cdot 10^{-4},
\ee
which is numerically very small, and should be close to the running of the spectral index $\alpha_s $ (\ref{ns2}).

\section{Basic Results}\label{conclusion} 
Our conclusion should be separated on two independent, but tightly related  parts.
The first portion represents some model independent very generic consequences of the $\qcd$-inflationary paradigm, while the second part represents very model dependent consequences of our proposal. 
 
We start with generic features of the $\qcd$ inflation.
As we formulated in section \ref{review}, the de Sitter behaviour (\ref{infl-EoS}) is very generic feature of the model which follows from three postulates formulated there.
It would be the final destination of our Universe if the interaction of the $\qcd$ fields with SM particles (\ref{coup})were always switched off.
This property  (\ref{infl-EoS}) is not related to the inflaton $\Phi$ or any other new propagating degrees of freedom; such fields do not exist in our framework.
Rather, this behaviour is a genuine quantum effect describing the dynamics of the topological sectors of the strongly coupled $\qcd$ as explained in details in \cite{Zhitnitsky:2013pna}. 

These features of this system are obviously very different from the conventional inflationary scenario normally formulated in terms of a scalar dynamical field $\Phi$, see the recent review papers \cite{Linde:2014nna,Brandenberger:2012uj} with opposite views on inflationary  cosmology.
For example, as is known, the initial value of the inflaton field $\Phi_{\rm in}$ (in the conventional scenario) must be larger than Plank scale to provide a sufficient number of e-folds $ N_{\rm Inf}\sim (\Phi_{\rm in}/M_{\rm PL})^2$.
A similar constraint is also required to support a slow-roll condition. 
In our framework, by contrast, the relevant $\qcd$ scale never exceeds the Planck mass (\ref{scale}), while the number of $e$-folds is determined by the gauge coupling constant (\ref{e-folds}).
Still, both mechanisms, the $\qcd$-inflation and conventional approach \cite{Linde:2014nna} eventually lead to the same de Sitter behaviour (\ref{infl-EoS}).
It would be very interesting to analyze and study the possible observational differences between these two fundamentally distinct  frameworks.

We listed above a number of very generic, model-independent consequences of the $\qcd$-inflationary paradigm.
We now want to mention some model dependent features of this paradigm.
These consequences have a very different status, as they are based on our specific assumptions about the evolution of helical instability.
We presented the corresponding results in sections \ref{EoS} and \ref{observations}.
The relevant solid technique which would make specific predictions about the EoS (and therefore on all spectral indices) is yet to be developed.
Nevertheless, the main point of this exercise is to argue that the numerical smallness of the corresponding indices (\ref{r1}),   (\ref{ns1}), (\ref{nt}), (\ref{ns2}), and (\ref{nt2}) is related to the numerical smallness of the gauge coupling constant.
The large number of $e$-folds expressed in terms of the same gauge coupling constant (\ref{e-folds}) is another manifestation of the same feature of the numerical smallness of the gauge coupling constant. 

We conclude this work (mainly devoted to inflation which is characterized by the Planck scale) with the following comment about a different field of physics with drastically different scales.
Namely, as we discussed at length in this paper, the  heart of the proposal is a fundamentally new type of energy which is not related to  any propagating degrees of freedom. 
Rather, this novel (non-dispersive) contribution to the energy has genuine quantum nature. The effect is formulated in terms of the tunnelling processes between topologically different but physically identical states.
This novel type of energy, in fact, has been well studied in the QCD lattice simulations, see footnote \ref{top} for references. 
Our comment relevant for the present study is that this fundamentally new type of energy can be, in principle, studied in a tabletop experiment by measuring some specific corrections to the Casimir vacuum energy  in the Maxwell theory as suggested in  \cite{Cao:2013na,Zhitnitsky:2013hba,Zhitnitsky_new}.
This fundamentally new contribution to the Casimir pressure emerges as a result of tunnelling processes, rather than due to the conventional fluctuations of the propagating photons with two physical polarizations.
This effect does not occur for the scalar field theory, in contrast with conventional Casimir effect which is operational for both: scalar as well as for Maxwell fields.
The extra energy computed in \cite{Cao:2013na,Zhitnitsky:2013hba,Zhitnitsky_new} is the direct analog of the ``strange energy'' which is the key player in the present work.
In fact, an extra contribution to the Casimir pressure emerges in this system as a result of nontrivial holonomy which can be enforced by  the nontrivial boundary conditions imposed in ref \cite{Cao:2013na,Zhitnitsky:2013hba,Zhitnitsky_new}.  

To conclude, we are not pretending to have solved a very complicated problem of inflation in the $\qcd$ framework as a large number of assumptions have been made along the way.
These assumptions obviously require further deep thinking and analysis.
Rather, the main goal of this work is to argue that the $\qcd$ inflationary paradigm is consistent with all presently available observations, including the recent BICEP2 discovery\cite{BICEP2}, which in fact, fixes the fundamental scale of the system: $\Lbar\simeq 10^{17}$ GeV. 
    
\section*{Acknowledgements}
I am thankful to David Gross  for  very useful discussions  we had  while  he was visiting  Vancouver in March 2014.  Our discussions on non-locality in QCD formulated in terms of large gauge transformation operator $\cal{T}$, on contact   term and its infrared origin  in QCD, on possible linear corrections $\sim H$ to the non-dispersive contact term in topological susceptibility, and many other deep and specific questions   are greatly appreciated. General comments by Andrei Linde are greatly appreciated. 

This research was supported in part by the Natural Sciences and Engineering Research Council of Canada.


\begin{thebibliography}{99}
\bibitem{BICEP2}
%%\cite{Ade:2014xna}
%\bibitem{Ade:2014xna} 
  P.~A.~R.~Ade {\it et al.}  [BICEP2 Collaboration],
  %``BICEP2 I: Detection Of B-mode Polarization at Degree Angular Scales,''
  arXiv:1403.3985 [astro-ph.CO].
  %%CITATION = ARXIV:1403.3985;%%
  %192 citations counted in INSPIRE as of 18 Apr 2014

\bibitem{inflation} A. Guth, Phys. Rev.  {\bf  D 23} (1981) 347;\\ A. Linde, Phys. Lett.  {\bf B 108} (1982) 389.
 %\bibitem{linde}
  %A. D. Linde, ÒInflationary Cosmology,Ó Lect. Notes Phys. {\bf 738}, 1 (2008) [arXiv:0705.0164 [hep-th]]
  
  
\bibitem{implications}

%\bibitem{Ma:2014vua} 
  Y.~-Z.~Ma and Y.~Wang,
  %``Reconstructing the Local Potential of Inflation with BICEP2 data,''
  arXiv:1403.4585 [astro-ph.CO].
  %%CITATION = ARXIV:1403.4585;%%

%\bibitem{Freese:2014nla} 
 K.~Freese and W.~H.~Kinney,
  %``Natural Inflation: Consistency with Cosmic Microwave Background Observations of Planck and BICEP2,''
  arXiv:1403.5277 [astro-ph.CO].
  %%CITATION = ARXIV:1403.5277;%%
  
  %\bibitem{Cheng:2014bma} 
   C.~Cheng and Q.~-G.~Huang,
  %``The Tilt of Primordial Gravitational Waves Spectra from BICEP2,''
  arXiv:1403.5463 [astro-ph.CO].
  %%CITATION = ARXIV:1403.5463;%%
  %10 citations counted in INSPIRE as of 18 Apr 2014
  
%\bibitem{Lyth:2014yya} 
 D.~H.~Lyth,
   %``BICEP2, the curvature perturbation and supersymmetry,''
  arXiv:1403.7323 [hep-ph].
  %%CITATION = ARXIV:1403.7323;%%

%\bibitem{Ho:2014xza} 
  C.~M.~Ho and S.~D.~H.~Hsu,
  %``Does the BICEP2 Observation of Cosmological Tensor Modes Imply an Era of Nearly Planckian Energy Densities?,''
  arXiv:1404.0745 [hep-ph].
  %%CITATION = ARXIV:1404.0745;%%
  %1 citations counted in INSPIRE as of 18 Apr 2014
  
  %\bibitem{Cheng:2014cja} 
  C.~Cheng and Q.~-G.~Huang,
  %``Constraint on inflation model from BICEP2 and WMAP 9-year data,''
  arXiv:1404.1230 [astro-ph.CO].
  %%CITATION = ARXIV:1404.1230;%%
  

%\cite{Zhitnitsky:2013pna}
\bibitem{Zhitnitsky:2013pna} 
  A.~R.~Zhitnitsky,
  %``Inflaton as an auxiliary topological field in a QCD-like system,''
  Phys.\ Rev.\ D {\bf 89}, 063529 (2014)
  [arXiv:1310.2258 [hep-th]].
  %%CITATION = ARXIV:1310.2258;%%
  
    %\cite{Zeldovich:1967gd}
\bibitem{Zeldovich:1967gd}
  Y.~B.~Zeldovich,
  %``Cosmological Constant and Elementary Particles,''
  JETP Lett.\  {\bf 6}, 316 (1967)
  [Pisma Zh.\ Eksp.\ Teor.\ Fiz.\  {\bf 6}, 883 (1967)].
  %%CITATION = ZFPRA,6,883;%%



   %\cite{Sola:2013gha}
\bibitem{Sola:2013gha} 
  J.~Sola,
  %``Cosmological constant and vacuum energy: old and new ideas,''
  J.\ Phys.\ Conf.\ Ser.\  {\bf 453}, 012015 (2013)
  [arXiv:1306.1527 [gr-qc]].
  %%CITATION = ARXIV:1306.1527;%%
  %20 citations counted in INSPIRE as of 28 Feb 2014

%\cite{Shuryak:2014gja}
\bibitem{Shuryak:2014gja} 
  E.~Shuryak,
  %``On Chiral Symmetry Breaking, Topology and Confinement,''
  arXiv:1401.2032 [nucl-th].
  %%CITATION = ARXIV:1401.2032;%%
  
%\cite{Yamamoto:2014vda}
\bibitem{Yamamoto:2014vda} 
  A.~Yamamoto,
  %``Lattice QCD in curved spacetimes,''
  arXiv:1405.6665 [hep-lat].
  %%CITATION = ARXIV:1405.6665;%%
  %1 citations counted in INSPIRE as of 10 Jul 2014



%\cite{Zhitnitsky:2013hs}
\bibitem{Zhitnitsky:2013hs} 
  A.~R.~Zhitnitsky,
  %``QCD as a topologically ordered system,''
  Annals Phys.\  {\bf 336}, 462 (2013)
  [arXiv:1301.7072 [hep-ph]].
  %%CITATION = ARXIV:1301.7072;%%
  %3 citations counted in INSPIRE as of 09 Aug 2013
  
  %\cite{Thomas:2011ee}
\bibitem{Thomas:2011ee} 
  E.~Thomas and A.~R.~Zhitnitsky,
  %``Topological Susceptibility and Contact Term in QCD. A Toy Model,''
  Phys.\ Rev.\ D {\bf 85}, 044039 (2012)
  [arXiv:1109.2608 [hep-th]].
  %%CITATION = ARXIV:1109.2608;%%
  %23 citations counted in INSPIRE as of 10 Jul 2014

%\cite{Frohlich:2002fg}
\bibitem{Frohlich:2002fg} 
  J.~Frohlich and B.~Pedrini,
  %``Axions, quantum mechanical pumping, and primeval magnetic fields,''
  cond-mat/0201236.
  %%CITATION = COND-MAT/0201236;%%
  
%\cite{Joyce:1997uy}
\bibitem{Joyce:1997uy} 
  M.~Joyce and M.~E.~Shaposhnikov,
  %``Primordial magnetic fields, right-handed electrons, and the Abelian anomaly,''
  Phys.\ Rev.\ Lett.\  {\bf 79}, 1193 (1997)
  [astro-ph/9703005].

%\cite{Akamatsu:2013pjd}
\bibitem{Akamatsu:2013pjd} 
  Y.~Akamatsu and N.~Yamamoto,
  %``Chiral Plasma Instabilities,''
  Phys.\ Rev.\ Lett.\  {\bf 111}, 052002 (2013)
  [arXiv:1302.2125 [nucl-th]].
  %%CITATION = ARXIV:1302.2125;%%
  %2 citations counted in INSPIRE as of 18 Aug 2013
  
  %\cite{Zhitnitsky:2012im}
\bibitem{Zhitnitsky:2012im} 
  A.~R.~Zhitnitsky,
  %``Local P Violation Effects and Thermalization in QCD: Views from Quantum Field Theory and Holography,''
  Nucl.\ Phys.\ A {\bf 886}, 17 (2012)
  [arXiv:1201.2665 [hep-ph]].
  %%CITATION = ARXIV:1201.2665;%%
  %11 citations counted in INSPIRE as of 15 Apr 2014
  
  %\cite{Zhitnitsky:2012ej}
\bibitem{Zhitnitsky:2012ej} 
  A.~R.~Zhitnitsky,
  %``P odd fluctuations and Long Range Order in Heavy Ion Collisions. Deformed QCD as a Toy Model,''
  Nucl.\ Phys.\ A {\bf 897}, 93 (2013)
  [arXiv:1208.2697 [hep-ph]].
  %%CITATION = ARXIV:1208.2697;%%
  %6 citations counted in INSPIRE as of 15 Apr 2014
  
   \bibitem{mukhanov}
V.~Mukhanov, {\it Physical Foundation of Cosmology}, Cambridge Univ.\ Pr.\ , 2005.
 
 \bibitem{planck}
 P.A.R. Ade et al [Planck Collaboration] ``Planck 2013 results.XXII. Constraints on inflation",
 arxiv:1305.5082[astro-ph.CO]
 
 %\cite{Smith:2014kka}
\bibitem{Smith:2014kka} 
  K.~M.~Smith, C.~Dvorkin, L.~Boyle, N.~Turok, M.~Halpern, G.~Hinshaw and B.~Gold,
  %``On quantifying and resolving the BICEP2/Planck tension over gravitational waves,''
  arXiv:1404.0373 [astro-ph.CO].
  %%CITATION = ARXIV:1404.0373;%%
  %8 citations counted in INSPIRE as of 18 Apr 2014
  
  %\cite{Linde:2014nna}
\bibitem{Linde:2014nna} 
  A.~Linde,
  %``Inflationary Cosmology after Planck 2013,''
  arXiv:1402.0526 [hep-th].
  %%CITATION = ARXIV:1402.0526;%%
  %14 citations counted in INSPIRE as of 27 Apr 2014
  
 %\cite{Brandenberger:2012uj}
\bibitem{Brandenberger:2012uj} 
  R.~Brandenberger,
  %``Do we have a Theory of Early Universe Cosmology?,''
  arXiv:1204.6108 [astro-ph.CO].
  %%CITATION = ARXIV:1204.6108;%%
  %5 citations counted in INSPIRE as of 05 Aug 2013

  %\cite{Cao:2013na}
\bibitem{Cao:2013na} 
  C.~Cao, M.~van Caspel and A.~R.~Zhitnitsky,
  %``Topological Casimir effect in Maxwell Electrodynamics on a Compact Manifold,''
  Phys.\ Rev.\ D {\bf 87}, 105012 (2013)
  [arXiv:1301.1706 [hep-th]].
  %%CITATION = ARXIV:1301.1706;%%
  %3 citations counted in INSPIRE as of 21 Aug 2013
  
  
%\cite{Zhitnitsky:2013hba}
\bibitem{Zhitnitsky:2013hba} 
  A.~R.~Zhitnitsky,
  %``Maxwell Theory on a Compact Manifold as a Topologically Ordered System,''
    Phys.\ Rev.\ D {\bf 88}, 105029 (2013)
  [arXiv:1308.1960 [hep-th]].
  
   
\bibitem{Zhitnitsky_new} 
  A.~R.~Zhitnitsky,
  %`` Topological Order  and Berry Connection for the Maxwell Vacuum on a four-torus,''
     [arXiv:1407.3804 [hep-th]].


  \end{thebibliography}
\end{document}